\documentclass [epsfig,manuscript,11pt,floatfix,amsmath,amssym,showpacs]{revtex4-1}
\usepackage[english]{babel}
\usepackage{graphicx}

\usepackage[dvips]{epsfig}

\usepackage{float}

\begin{document}

\newcommand{\xv}{{\mathbf x}}
\newcommand{\qv}{{\mathbf q}}
\newcommand{\av}{{\mathbf a}}
\newcommand{\cH}{{\cal H}}
\newcommand{\lB}{\ell_B}
\newcommand{\apa}{a_\parallel}
\newcommand{\ape}{a_\perp}
\newcommand{\Sv}{\hat{{\bf S}}}
\newcommand{\ev}{\hat{{\bf e}}}
\newcommand{\tv}{\hat{{\bf t}}}
\title{Intradomain Textures in Block Copolymers:  Multizone Alignment and Biaxiality  }
\author{Ishan Prasad}
\affiliation{Department of Chemical Engineering, University of Massachusetts, Amherst, Massachusetts 01003, USA}
\author{Youngmi Seo}
\affiliation{William G. Lowrie Department of Chemical and Biomolecular Engineering, The Ohio State University, Columbus, Ohio 43210, USA}
\author{Lisa M. Hall}
\affiliation{William G. Lowrie Department of Chemical and Biomolecular Engineering, The Ohio State University, Columbus, Ohio 43210, USA}
\author{Gregory M. Grason}
\affiliation{Department of Polymer Science and Engineering, University of Massachusetts, Amherst, MA 01003, USA}

\begin{abstract}
Block copolymer (BCP) melt assembly has been the subject of decades of study, with  focus largely on self-organized spatial patterns of periodically-ordered segment density.  In this study, we demonstrate that underlying these otherwise well-known composition profiles (i.e. ordered lamella, cylinders, spheres and networks) are generic and heterogeneous patterns of segment orientation that couple strongly to morphology, even in the absence of specific factors that promote intra- or inter-chain segment alignment.  We employ a combination of self-consistent field theory and coarse-grained simulation methods to measure polar and nematic order parameters of segments in a freely-jointed chain model of diblock melts.  We show that BCP morphologies are generically characterized by a {\it multizone} texture, with segments predominantly aligned normal and parallel to inter-domain interfaces in the respective brush and interfacial regions of the microdomain.  Further, morphologies with anisotropically-curved interfaces (i.e. cylinders and networks) exhibit biaxial order that is aligned to the principal curvature axes of the interface.  Understanding these generic features of intra-domain texture provide new opportunities for manipulating multi-scale structure and functional properties of BCP assembly.
\end{abstract}

\maketitle
Block copolymer (BCP) melts assemble into a rich array of periodic morphologies such as spheres, cylinders, networks, and layers~\cite{Fredrickson1996} depending on the composition, architecture, and interactions between blocks~\cite{Grason2006,abetz,matsen_2012}. Over the past several decades, investigations of BCP assembly have focused on equilibrium composition profiles of chemical segments $\phi_{\alpha}(\textbf{x})$ (for component $\alpha$) and their connection to copolymer architecture~\cite{Matsen2002}. Periodically-ordered morphologies break not only continuous translational symmetries of a disordered melt, but also its continuous rotational symmetry.  As a consequence, BCP morphologies necessarily possess orientational order, both at the scale of micro-domain lattice and at a sub-domain scale. Despite the extensive study of the spatially-ordered composition profiles of BCP and their now widespread applications in nanotechnology, knowledge of orientational order of chain segments that generically underlies these spatial patterns is conspicuously lacking.

In this paper we use self-consistent field (SCF) theory and coarse-grained simulations to analyze intra-domain segment orientation patterns in BCP melts to understand i) in which directions constituent segments are aligned within microphase domains; and ii) how alignment varies with BCP morphology. Our analysis is based on the simplest models of flexible BCPs, which lack explicit orientational interaction between segments, nonetheless exhibit generic patterns, or textures, of orientational order. Naively, our questions are akin to those addressed in studies of liquid crystalline textures confined in volumes of differing size, shape, and topology, from droplets~\cite{lavrentovich, fernandez} to 3D periodic networks \cite{Serra2011}. In such systems, it is well known that textures within micro-/nano-confined volumes are highly dependent on the shape of the confining volume, on orientational symmetries of the ordered phases~\cite{sec, depablo}, and, crucially, on the anchoring effect of alignment on the confining surface~\cite{lavrentovich_kleman}. Analogous alignment may be expected from the spontaneously formed interface between unlike components in BCPs, in which case one may ask, are segments aligned parallel (planar anchoring) or normal (homeotropic anchoring) to inter-domain surfaces in BCP assemblies? Curiously, SCF studies of the nematic order parameter in phase-separated mixtures of homopolymers show a generic tendency of segmental alignment parallel to the interface over the interfacial width \cite{Szleifer1989, Carton1990, Morse}, while the SCF prediction of the polar order parameter in BCP micro domains shows instead a normal alignment more characteristic of a SmA-like order~\cite{Zhao2012a, Chen2013}. Here, we show that {\it both} normal and parallel segment alignment coexist generically within subdomains of BCP, albeit in different regions of the microstructure. We describe basic principles that control relative strengths and directionality alignment in different BCP morphologies and in different subdomains of a given morphology. Perhaps most surprising, we report the generic emergence of biaxial segment order in morphologies with anisotropically-curved interfaces.

We consider a freely-jointed chain model of a diblock copolymer melt~\cite{doiedwards}, where chains possess $N_A$ and $N_B$ Kuhn segments of $A$- and $B$-type monomers, respectively.  For simplicity, we present the case of equal segment length $a$ and volume $\rho_0^{-1}$.  Chains are labeled by segment number $n$, which runs from $0<n \leq N_A$ in the $A$-block and $N_A<n\leq N_A+N_B=N$ in the $B$-block.  In the mean-field, or self-consistent field (SCF) approximation, chain conformations are encoded in end-distribution functions, $q^+(n, \xv)$ and $q^-(n, \xv)$, which describe the statistical weights of disjoint sections of the chain from the respective $A$ ($n=0$) and $B$ ($n=N$) ends to reach $\xv$ at the $n$th segment.  Thus, the probability (per unit volume) of the $n$th segment of the diblock at $\xv$ is $\rho(\xv,n) = q^+(n,\xv) q^-(n, \xv)/Z$, where $Z$ is the single-chain partition function $Z = \int d^3\xv ~q^+(n,\xv) q^-(n, \xv)$.  The mean-field scalar order parameters, volume fractions of $A$ and $B$, follow directly from the end-distributions
\begin{equation} \label{eq:OPphi}
\phi_{\alpha} (\xv) = \Big \langle \rho_0^{-1} \sum_{\nu \in \alpha} \delta (\xv_\nu - \xv) \Big \rangle= \frac{ V}{N} \int_{\alpha} dn ~ \rho(\xv,n) ,
\end{equation}
where $\nu \in \alpha$ labels all segments of type $\alpha$ in the melt, $\alpha=A$ or $B$.  Due to random-walk chain conformations~\footnote{Here, we take the limit $N \to \infty$ and $a \to 0$ with$N^{1/2} a$, such that large-scale structure is described by Gaussian walk statistics.}, end distributions obey the modified diffusion equation~\cite{helfand, Matsen2002, fredrickson2006equilibrium},
\begin{equation} \label{eq:MDE}
\pm \frac{ \partial q^\pm}{\partial n} = \frac{a^2}{6} \nabla^2 q^\pm - w(n, \xv) q^\pm ,
\end{equation}
where $w(n, \xv) =  \Theta(N_A-n) w_A(\xv)+ \Theta(n-N_A) w_B(\xv)$, with $w_{A/B}(\xv)$ the spatially-varying chemical potential field for $A$ or $B$ generated by local segment (scalar) interactions~\footnote{End-distributions satisfy spatially uniform initial conditions, $q^+(n=0) =  q^-(n=N) =1$.}.  The chemical potential fields satisfy the mean-field, or self-consistency, condition, $w_{A/B}(\xv) = \chi \phi_{B/A}(\xv) + \xi(\xv),$ where Flory parameter, $\chi$, describes repulsive interactions between unlike species and the pressure field, $\xi(\xv)$, acts on both species to maintain constant density (i.e. $\phi_A (\xv)+ \phi_B(\xv) = 1$). Equilibrium ordered states are determined by solving eq. \ref{eq:MDE} for spatially periodic patterns of $\phi_A (\xv)$ and $\phi_B (\xv)$, optimized with respect to symmetry and unit cell dimensions \cite{fredrickson2006equilibrium}.  We employ the PSCF code~\cite{Arora} to compute end-distributions for ordered morphologies at fixed segregation strengths $\chi N$ and chain composition $f=N_A/N$.

\begin{figure}
\includegraphics[width=0.6\textwidth]{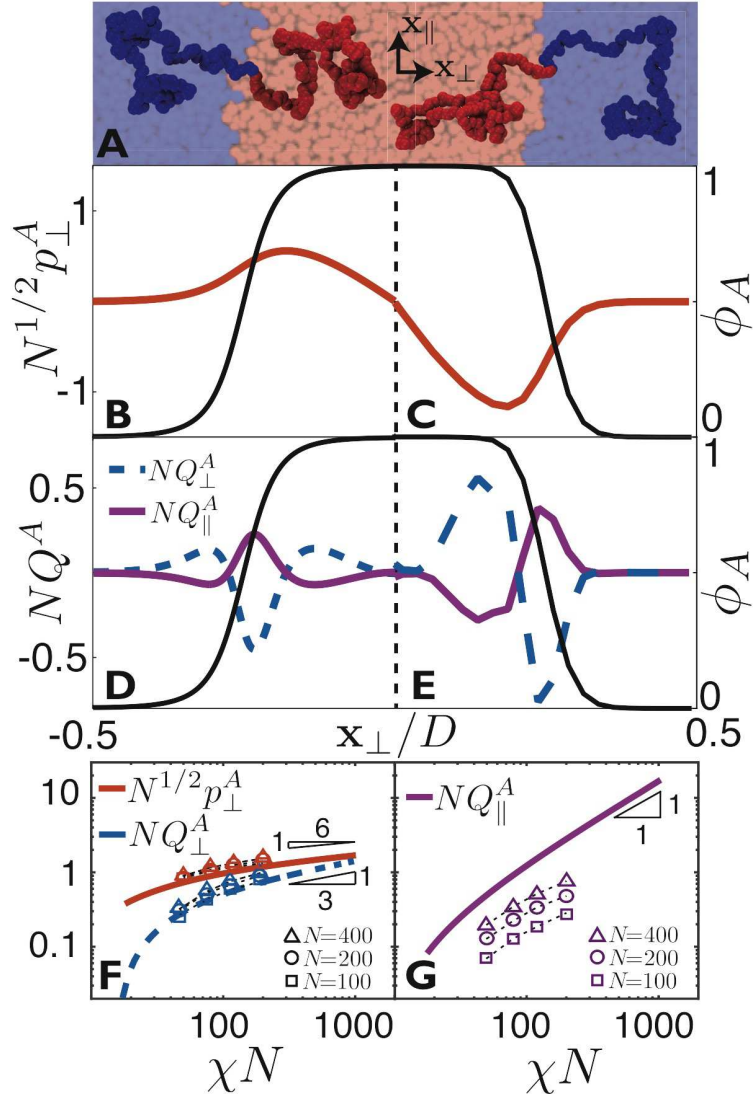}
\caption{ (A) MD simulation snapshot showing lamellar domains, where $\xv_{\perp}$ and $\xv_{\parallel}$ mark respective normal and tangential directions. (B)-(E) Order parameter (left y-axis) and volume fraction (black curve, right y-axis) profiles for A-block segments in $f=0.5$ lamella, with (B), (D) showing SCF results ($\chi N =30$) and (C), (E) showing MD results ($\chi N =80$) .  (B-C) show the normal component of ${\bf p}_A$ (parallel component is 0), and (D-E) show normal and parallel components of $Q_{ij}^A$. Peak values of normal (F) and parallel (G) components of polar and nematic order in $f=0.5$ lamella are plotted vs. $\chi N$, with SCF results shown as curves and MD results as open symbols. }
\label{fig1}
\end{figure}  

Orientational order of segments in block copolymer morphologies can be described by two types of order parameters, both encoded in spatial derivatives of end-distribution functions.  A {\it polar} order parameter ${\bf p}_{\alpha}(\xv)$ tracks the vectorial orientation of segments~\cite{Zhao2012a}, since $A$ and $B$ ends are distinguishable.  Assigning $\hat{r}_\alpha$ to orientation of segment $\alpha$ (directed from $A$ to $B$ end),
\begin{equation} \label{eq:OPp}
{\bf p}_{\alpha}(\xv) = \Big \langle \rho_0^{-1} \sum_{\nu \in \alpha} \hat{r}_\alpha ~ \delta (\xv_\nu - \xv) \Big \rangle= \frac{ V}{N} \int_{\alpha} dn ~ {\bf J} (\xv,n) ,
\end{equation} 
where the segment flux  is given by ${\bf J} =a(q^+ \nabla q^- - q^- \nabla q^+)/(6Z)$.   A {\it nematic} order parameter ${\bf Q}^{\alpha} (\xv)$, a symmetric, traceless tensor, tracks anisotropy of segments consistent with head-tail symmetry (or $\hat{r}^\alpha \to - \hat{r}^\alpha$) of alignment~\cite{de1993physics} (where $i,j,k$ are spatial indices),
\begin{equation} \label{eq:OPQ}
Q_{ij}^{\alpha} (\xv) = \Big \langle \rho_0^{-1} \sum_{\nu \in \alpha}\Big[ (\hat{r}^\alpha)_i (\hat{r}^\alpha)_j - \frac{\delta_{ij}}{3} \Big] \delta (\xv^\alpha - \xv) \Big \rangle
=  \frac{ V}{N} \int_{\alpha} dn ~ \big[ \mathcal{J}_{ij} (\xv,n) - \frac{\delta_{ij} }{3}  \mathcal{J}_{kk} (\xv,n)\big] ,
\end{equation}
where $\mathcal{J}_{ij}= a^2(q^+ \partial_i \partial_j q^- + q^- \partial_i \partial_j q^+ - \partial_i  q^+ \partial_j  q^- - \partial_i  q^- \partial_j  q^+)$/(60Z).  To compare to our (mean field) SCF results, we perform molecular dynamics (MD) simulations of analogous freely-jointed bead-spring chains, specifically using finitely extensible nonlinear elastic (FENE) bonds and the repulsive part of the Lennard-Jones (LJ) potential for all pairwise interactions~\cite{grest}. Simulations do not rely on the mean-field approximation and capture inter-segment correlation effects absent in the SCF model. Phase separation is driven by increased A-B repulsion strength $\epsilon_{AB}$, mapped to $\chi_{AB}$ as in ref. \cite{grest} (see SI for details). We use a Langevin thermostat and Nos\'{e}-Hoover barostat with pressure $5\epsilon \sigma^{-3}$ as is typical in Kremer-Grest simulations, initialized in an already microphase-separated structure, and allow the box shape to adjust to reach the proper domain spacing~\cite{seo}. Vector and tensor order parameters are computed from bond-vector, $\hat{r}_\alpha$, distribution extracted from equilibrated configurations. Data is binned by distance from the lamellae center of mass, or, for the cylinder morphology, by radial distance from the center of mass of each cylinder (see Appendix D for details.)

To demonstrate characteristic ``multizone" features of segment orientation we first analyze lamellar order.  Fig. 1B-E show profiles of polar and nematic order of $A$ block segments for well-segregated lamella at $\chi N = 30$, $f=0.5$. Turning first to the ``brush" zone deep in the $A$-rich domain (i.e. where $\phi_A \approx 1$) , we find the intuitive result of {\it normal} segment orientation (i.e. SmA-like).  Defining $\parallel$ and $\perp$ directions relative to $AB$ interface, symmetry along the layer guarantees $p_\parallel^A=0$, while the profile of  $p_\perp^A$ is odd with respect to the $A$-domain center, highlighting outward splay of chains away from the bilayer interfaces.  In this region, the nematic order shows similar uniaxial normal alignment with $Q^A_\perp> 0$ consistent with chain stretching in brush-like domains away from the $AB$ interface. The degree of alignment can be estimated by a simple Langevin model of chains subject to a mean-field tension, $\tau \approx k_B T D/N a^2$, which represents the self-consistent effect of inter-segment pressure holding free ends a distance proportional to the domain width $D$ from the interface.  Assuming the probability of angle $\theta$ with respect to the normal, $P(\theta) \approx \sin \theta ~ e^{\tau a \cos \theta/k_B T}$, the respective magnitudes of polar and nematic normal order are $ p^\alpha_\perp \approx D_\alpha/(N_\alpha a)\sim N^{-1/2} (\chi N)^{1/6}$ and  $Q^\alpha_\perp \approx (D_\alpha)^2/(N_\alpha a)^2 \sim N^{-1} (\chi N)^{1/3}$, where strong-segregation theory gives $D_\alpha \sim D \approx (\chi N)^{1/6} N^{1/2}a$~\cite{Semenov1985} which captures asymptotic $\chi N \gg 1$ power laws observed for peak SCF order parameters in Fig. 1F.

\begin{figure}
\includegraphics[width=0.6\textwidth]{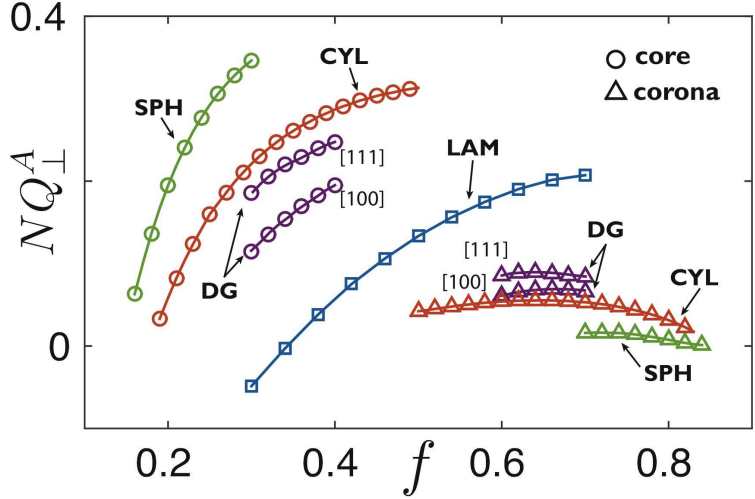}
\caption{Peak values of normal component of nematic OP for A segments, $Q_\perp^A$ in BCC spheres (SPH), hexagonal cylinders (CYL), cubic double-gyroid (DG), and lamellar (LAM) phases. Core (corona) reflects location of A block on inner (outer) side of the AB interface.  Max. and min. values along both [111] and [100] axes are shown for DG (see Appendix C).}
\label{fig2}
\end{figure}  

Turning now to the interfacial zone ($\phi_A \approx \phi_B = 1/2$), the nematic order parameter in Fig.~\ref{fig1}D,E reveals that segment alignment becomes {\it tangential} (i.e. $Q^\alpha_\parallel >0$ and $Q^\alpha_\perp<0$) near the  inter-domain boundary, implying that both {\it normal} and {\it tangential} segment alignment coexist within block copolymer domains, albeit at different spatial regions.  The existence of tangential alignment at the interface, though arguably less intuitive than normal ordering in the brush, is nonetheless a generic feature of the statistics of random-walks at a composition boundary, even in the absence of physical interactions that promote (inter-/intra-chain) segment alignment.  This follows from the fact that segment configurations spanning from the ``rich" to ``poor" side of an interface are depleted relative those extending along the boundary, leading to a net bias of tangential orientations analogous to the effect of ``hard wall"~\footnote{More explicitly, the distribution of interfacial orientation follows directly as the probability $\rho(\xv, \hat{r})$ of segment orientation $\hat{r}$ at $\xv$ that {\it any two} (like) chain sections are span from $\xv-a\hat{r}/2$ to $\xv+a\hat{r}/2$.  In other words, the probability $\rho(\xv, \hat{r})$ is the {\it geometric mean} of segment density at nearby points, $\rho(\xv, \hat{r}) \simeq \sqrt{ \phi(\xv-a \hat{r}/2)} \times \sqrt{ \phi(\xv+a \hat{r}/2)} $.  Because $\ln \phi(\xv)$ necessarily becomes non-convex at the interface (along the normal direction, {\bf N}), fewer orientations span perpendicular ($\perp$) than parallel ($\parallel$) to the AB interface, $\rho_\perp (\xv)-\rho_\parallel(\xv)\simeq  (a^2/8) \phi ~ \partial_{{\bf N}}^2 \ln \phi  <0$.}. The degree of orientational order is directly related to the interfacial structure. End-distribution functions become independent of $n$ near a well-segregated interface as ends and junction points are rare. According to eq. \ref{eq:OPphi} the segment distributions $q^+$ and $q^-$ become approximately $\propto \sqrt{\phi_\alpha (\xv)}$~\cite{helfand_wasserman, Goveas}.  Inserting this assumption into eq. (\ref{eq:OPQ}) leads directly to the general form of the nematic order parameter near a sharp interface,  
\begin{equation} \label{eq:Qint}
Q^{\alpha}_{ij} ({\rm inter.}) = \frac{a^2}{60} \phi _{ \alpha} \big[\partial_i \partial_j  
\ln  \phi _{ \alpha}- \frac{\delta_{ij}}{3}  \nabla^2 \ln \phi _{ \alpha}\big] ,
\end{equation}
For flat lamellar domain interfaces, nematic order along the layer normal ${\bf N}$ takes the simple form $Q^{\alpha}_{\perp} ({\rm inter.}) \simeq( a^2/90) \phi_\alpha \partial_{N}^2 \ln \phi_\alpha$, where $\partial_N = ( {\bf N} \cdot \nabla) $. As width of the interface has the form $\Delta= a (2/3 \chi)^{1/2}$~\cite{Semenov1985}, the strength of tangential interfacial alignment is {\it independent of $N$}, scaling as $Q^\alpha_\perp \sim \chi=-2Q^\alpha_\parallel \sim -\chi$ for sharp (large $\chi N$) and generically non-convex interfaces, confirmed for peak interfacial order in lamella in Fig.~\ref{fig1}G.  

Profiles of polar and nematic order parameters from MD simulations of lamellar morphologies are shown in Fig.~\ref{fig1}C,E clearly exhibiting generic features of the ``multi-zone pattern" predicted by SCF theory.  This is despite key microscopic differences between the simulation and theory, including i) local packing constraints of finite-sized spherical monomers in the simulation absent from the SCF model, ii) the ability to vary local density  in simulations (the system can reduce unfavorable interactions by lowering interfacial density), while SCF enforces constant local density, and iii) the fact that MD simulations use relatively low $N=100-400$ for tractability, requiring relatively large $\chi$ for strong segregation, while the SCF considers the limit of $\chi \ll 1$ (or $N \to \infty$). Fig.~\ref{fig1}F,G show SCF and MD predictions are in closer agreement for normal ordering in brush than for tangential alignment at the interface, where bead-spring simulations show a weaker alignment, presumably related to the fact that $a \gtrsim \Delta$ for sufficiently large $\chi$.  However, we find that as $N$ increases (at fixed $\chi N$), the segment alignment in simulations tends towards SCF predictions, consistent with the approach towards $N \to \infty$.  

\begin{figure}
\includegraphics[width=0.65\textwidth]{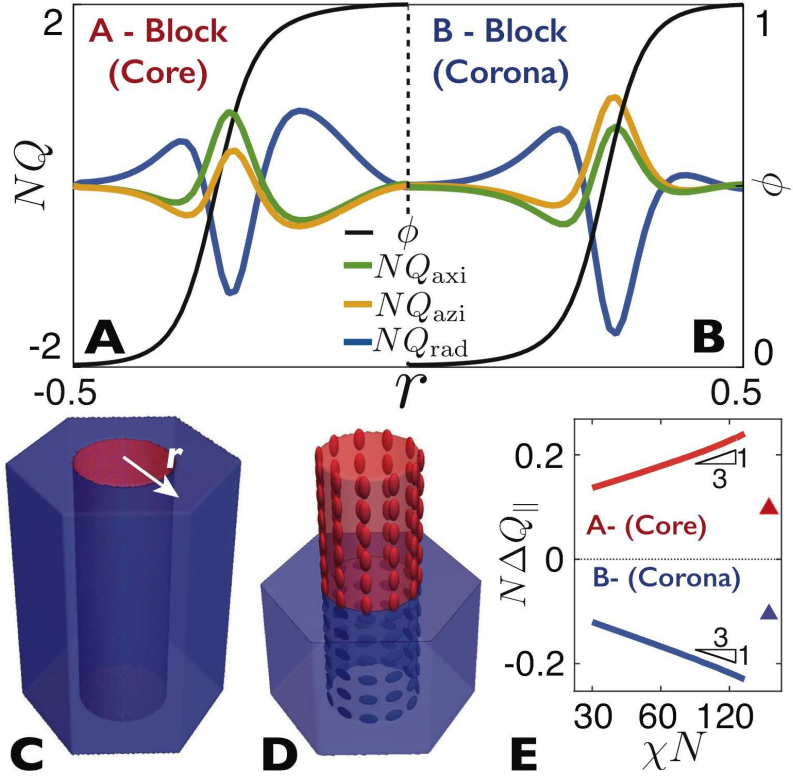}
\caption{(A) and (B) show nematic profiles for A-(core) and B-block (coronal) segments, respectively, in the CYL phase at $\chi N = 30$ and $f=0.3$.  $Q_{axi}$, $Q_{azi}$ and $Q_{rad}$, label the respective  axial, azimuthal and radial components, defined with respect to the center axis of the cylinder with radial distance $r$ shown in schematic (C).  In (D), red and blue ellipsoids illustrate the biaxial interfacial order for core (A-block) and coronal (B-block) segments, respectively, where the dimension of axes reflect magnitude of $Q^\alpha_{ij}$.  In (E), peak values of $\Delta Q_{\parallel} \equiv Q_{\mathrm{axi}} - Q_{\mathrm{azi}}$ are plotted for SCF results (solid lines) and MD simulations (triangles) at $f=0.25$ for $N=100$. }
\label{cyl}
\end{figure}  

Moving beyond the simplest case of lamella, the geometry of equilibrium morphology has strong effects on the respective normal and tangential alignment zones. As block composition becomes increasingly asymmetric (i.e. as $|f - 1/2|$ grows), minority block domains tend to form on the inside of $AB$ interface of increasing inward curvature~\cite{matsen_bates}.  As chains occupy a fixed volume, simple geometric arguments imply that increasing interfacial curvature leads to a tendency to {\it relax} the outer block length at the expense of {\it extending} the block on the inward curvature side~\cite{olmsted_milner}.  Accordingly, normal order increases with a power of $D_\alpha$, such that in minority (majority) subdomains, normal alignment of in brushes increases (decreases) with increasing inter-domain curvature from lamella $\to$ double-gyroid $\to$ cylinders $\to$ spheres, consistent with the variation of peak $Q^A_{\perp}$ across the composition range in Fig.~\ref{fig2}.

The interface shape has arguably a more profound effect on alignment in the interfacial zone. Consistent with arguments above, the normal component of $Q^\alpha_{ij}$ is generically negative at the $AB$ interface, indicating net positive alignment in the parallel, in-plane directions.  However, for morphologies with anisotropically curved interfaces (i.e. cylinders and tubular networks), in-plane alignment couples to principle curvature axes as illustrated by the nematic order profile of a cylinder morphology in Fig.~\ref{cyl}. This coupling follows from the nematic order parameter, $Q^\alpha_{IJ} \equiv {\bf e}_I \cdot {\bf Q}^\alpha \cdot  {\bf e}_J$, projected onto an orthonormal basis $\{ {\bf e}_1,  {\bf e}_2, {\bf e}_3={\bf N} \}$ that is aligned to the local normal (${\bf N}$) and tangent plane (spanned by ${\bf e}_1$ and ${\bf e}_2$).  Using eq.\ref{eq:Qint} and the fact that domain interfaces are isolevels of volume fraction with $\nabla \phi_\alpha = (\partial_N \phi_\alpha) {\bf N}$, the in-plane nematic order in the interfacial zone is (see Appendix B),
\begin{equation} \label{eq:curv}
Q^{\alpha}_{IJ} ({\rm inter.})  \simeq \frac{a^2}{60} \Big[-\partial_N \phi_\alpha C_{IJ} -\frac{\delta_{IJ}}{3} \big(\partial^2_N \phi_\alpha -|\partial_N \phi_\alpha|^2/\phi_\alpha-2 H \partial_N \phi_\alpha \big) \Big] \ \ \ \ {\rm for} \ I,J = 1,2 ,
\end{equation}
where $C_{IJ} = {\bf N} \cdot \big[ ({\bf e}_I \cdot \nabla) {\bf e}_J\big]$ is the curvature tensor of the interface and $H = (C_{11} +C_{22})/2$ is the mean curvature~\cite{kamien}.   For anisotropic-curved interfaces, where $C_{IJ} \neq H \delta_{IJ}$, the in-plane segment order parameter is also anisotropic, with maximal alignment along either the direction of maximal or minimal curvature. Taking ${\bf e}_1$ and ${\bf e}_2$ to be principal curvature directions, we measure in-plane alignment anisotropy $\Delta Q_{\parallel} \equiv Q_{11}-Q_{22}$, and find $\Delta Q_{\parallel} \simeq - \frac{a^2}{60Z} (\kappa_1-\kappa_2)\partial_N \phi_\alpha$, where $\kappa_1$ and $\kappa_2$ are principal curvatures. Surface curvature falls with domain size as $\kappa \propto D^{-1}$, while $|\partial_N \phi_\alpha| \sim \Delta^{-1}$, and interfacial anisotropy grows with $N \Delta Q_\parallel \sim (\chi N)^{1/3}$ in the strong-segregation limit (shown in Fig.~\ref{cyl}E). For cylindrical domains of core radius $R_c$, where $\kappa_1=0$ and $\kappa_2=-1/R_c$ (taking ${\bf N}$ to be outward), $\partial_N \phi_\alpha$ switches sign from negative when $\alpha$ is the inner domain to positive when it is the outer domain. Hence, not only are interfacial segments aligned to the local curvature directions, but this alignment along principal eigenvectors of ${\bf Q}$ is distinct for core- vs. coronal-block segments at that interface.  Peak values of nematic segment order in cylinder phases show that core-block segments most strongly align with the {\it axial} direction, while coronal-block segments (at that same interface) align most strongly to the {\it coaxial} direction (shown schematically in Fig.~\ref{cyl}D).

\begin{figure*}
\includegraphics[width=0.95\textwidth]{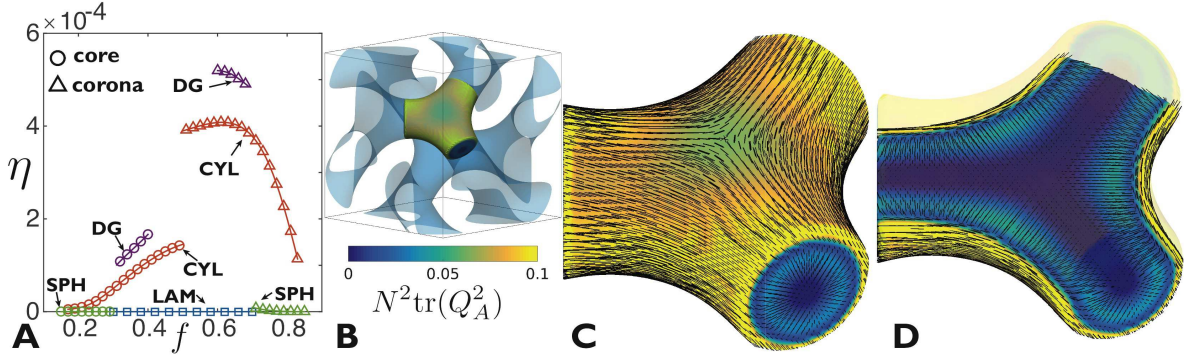}
\caption{(A) Plot of volume-averaged biaxiality, $\eta_A \equiv \big[ \mathrm{tr}({\bf Q}_\alpha^2)\big]^2 - 54\big[\mathrm{det}({\bf Q}_\alpha)\big]^2$, as a function of composition $f$ for different diblock morphologies at $\chi N = 30$. (B-D) show the 3D nematic director field of the tubular minor domain of a DG network at $f=0.33$, in the 3-fold region highlighted in (B).  (C) shows the director profile at the AB interface ($\phi =1/2$), while (D) shows the profile through a core section. }
\label{gyroid}
\end{figure*}

We note further that according to eq.(\ref{eq:curv}), tangential ordering at anisotropic interfaces are marked by {\it biaxial segment order}, with three unequal eigenvalues of ${\bf Q}^\alpha$ (one negative and two unequal positive values roughly aligned to the normal and principal curvature directions of the interface, respectively). Adopting methods developed to describe biaxial phases of liquid crystals, we quantify the degree of biaxiality \cite{Freiser1970, Alben1973} from the rotational invariant, $\eta_\alpha \equiv \big[ \mathrm{tr}({\bf Q}_\alpha^2)\big]^2 - 54\big[\mathrm{det}({\bf Q}_\alpha)\big]^2$, that increases from zero as eigenvalues of ${\bf Q}$ become unequal.  In Fig.~\ref{gyroid}A, we show biaxiality (volume averaged) of A segments, $\langle \eta_A \rangle$, in competing diblock phases (at $\chi N = 30$), indicating that biaxial order is absent (present) for phases with isotropic (anisotropic) in-plane curvature.
Negative Gaussian curvature of network interfaces~\cite{olmsted_milner} implies larger curvature anisotropy ($\kappa_1-\kappa_2$), and hence, the largest segment biaxiality. Fig.~\ref{gyroid}B-D, shows the complex pattern of nematic order (as illustrated by the director field) in minor, tubular domains of the double-gyroid. Notably, alignment in core brush and interfacial zones implies the formation of point and line defects~\cite{Alexander2012}.   The director exhibits a hedgehog defect meeting at the 3-fold junction of three $+1$ disclination-like lines that thread along center of the tubular domains.  At the interface, locking of the director to the curvature axes leads to the formation of two $-1/2$ disclinations on antipodal points of the three-fold junction that localize the conflict with in-plane order and Gaussian curvature of the interface. 

To conclude, while the degree of normal alignment of brush segments increases inversely with chain length ($\sim N^{-1/3}$ and $\sim N^{-2/3}$ for polar and nematic order, respectively), we find that tangential alignment at the interface is independent of chain length, grows in proportion to $\chi$.  This suggests that even in flexible diblocks, tangential alignment will approach significant levels ($Q_{ij} \sim 1$) in high-$\chi$ systems~\cite{Gopalan, ellison}. This generic orientational pattern of flexible chain diblocks is also a necessary reference point for studying orientational order in systems with chains described by additional tendencies promoting intra-chain (i.e. persistence) and inter-chain segmental alignment~\cite{matsen_96, duchs, Zhao2012a, Chen2013, Chen2016}.  For example, recent studies of BCPs with chiral polymer blocks~\cite{ho} suggest these systems may be described by an additional preference of twisted (e.g. cholesteric) packing in the chiral micro domains~\cite{Zhao2013, Wang2013, Grason2015}, a pattern of gradient orientation that competes with the entropically favorable multi-zone alignment described here. Finally, we note that prediction of segment alignment at anisotropic inter-domain surfaces may have key, as of yet, unexplored consequences for behavior of functional BCP assemblies.  For example, one expects that the performance of hybrid BCP materials where functionality emerges from the interface of cylinder mesophase and relies on directional processes (e.g. photo/optical response, charge transport) will  exhibit a strong dependence on core vs. coronal placement of functional blocks.

\begin{acknowledgments}
We are indebted to A. Arora for key input regarding the implementation of PSCF, and to E. Thomas and R. Kamien for valuable discussions.  This work was supported by U.S. Department of Energy, Office of Science, Office of Basic Energy Sciences, under Award Number DE-SC0014549 (SCF theory, IP and GG) and National Science Foundation under Grant No. 1454343 (MD simulations, YS and LMH), using computational facilities at the Massachusetts Green High Performance Computing Center and the Ohio Supercomputer Center.

\end{acknowledgments}

\appendix

\section{Orientational Order Parameters}

\label{OPs}

Here, we will briefly derive the relationship between segment order parameters, both vectorial $\mathrm{\textbf{p(x)}}$ and tensorial $\mathrm{\textbf{Q(x)}}$, and segment-distribution functions, $q^{\pm}(\xv)$, based on the freely-jointed chain model for a polymer chain, subject to the spatially varying chemical potential $w(n,\xv)$ for type-$\alpha$ segments.  The transfer probability of the $n$th segment of the chainfrom $\mathrm{\textbf{x}}$ to $\mathrm{\textbf{x}'}$ is
\begin{equation} \label{Seq:weight}
s (n; \textbf{x},\textbf{x}') = M^{-1} e^{- w(n,\bar{{\bf x}}) } \delta \big(|\xv-\xv'|-a\big)
\end{equation}
where we take the chemical potential to act on the mid-point $\bar{\xv} = (\xv+\xv')/2$ of the segment and $M= 4 \pi a^2$ normalizes the probability in the absence of field. Here, we explicitly consider the limit of $N \gg 1$ such that segment size $a$ becomes vanishingly small for fixed mean-square random chain size, $aN^{1/2}$.  Similarly, this assume $\chi \ll 1$ such that $\chi N$ is finite. The probability distribution $P(n;\xv,\hat{r} )$ of the $n$th segment of the chain at $\xv$ with orientation $\hat{r}$ is then,
\begin{multline}
P(n;\xv,\hat{r}) = \frac{1}{ M {\cal Z}} q^+(n-1,\xv-\hat{r}a/2) q^-(n,\xv+\hat{r}a/2)e^{- w(n,{\bf x}) } \\
 \simeq \frac{1}{ {\cal Z}}\Big\{ q^+(n, \xv) q^-(n,\xv) + \frac{a}{2} \hat{r}_i \big[ q^+(n,\xv) \partial_i q^-(n, \xv) - q^-(n,\xv)  \partial_i q^+(n, \xv) \big]\\
 + \frac{a^2}{8} \hat{r}_i \hat{r}_j\big[ q^+(n,\xv) \partial_i \partial_j q^-(n,\xv) +q^-(n,\xv) \partial_i \partial_j q^+(n,\xv) -2 \partial_iq^+(n,\xv)  \partial_j q^-(n,\xv)\big] \Big\}
 \end{multline}
where we retain terms up to second order in $\hat{r}$ and drop contributions proportional to $w(\xv)$ and $\partial q/\partial n$ which are of order $\chi \ll 1$.  

The polar order parameter is simply the first moment of this distribution at $\xv$ averaged over all $\alpha$ segments,
\begin{equation}
p_i^\alpha (\xv) = \frac{V}{N} \int_{n\in \alpha} dn ~ \int d^2 \hat{r}~\hat{r}_i P(n;\xv,\hat{r}) = \frac{a V}{6 N} \int_{n\in \alpha} dn ~ (q^+ \partial_i q^- - q^- \partial_i q^+ ) ,
\end{equation}
where we used $\int d^2 \hat{r} ~\hat{r}_i \hat{r}_j = \delta_{ij}/3$.  Likewise, the nematic order parameter tensor follows from the second moment (the traceless, symmetric part),
\begin{multline}
Q_{ij}^\alpha (\xv) = \frac{V}{N} \int_{n\in \alpha} dn ~ \int d^2 \hat{r}~\Big( \hat{r}_i  \hat{r}_j - \frac{\delta_{ij} }{3} \Big) P(n;\xv,\hat{r}) \\
= \frac{ a^2 V}{60 N} \int_{n\in \alpha} dn ~\Big(q^+ \partial_i \partial_j q^- +q^- \partial_i \partial_j q^+ -\partial_i q^+  \partial_j q^- - \partial_i q^-  \partial_j q^+ \\  - \frac{\delta_{ij} }{3} \big[q^+ \nabla^2 q^- +q^- \nabla^2 q^+-2 (\nabla q^+) \cdot(\nabla q^-) \big] \Big)
\end{multline}
where we used additionally $\int d^2 \hat{r} ~\hat{r}_i \hat{r}_j \hat{r}_k \hat{r}_\ell = \frac{1}{15}( \delta_{ij} \delta_{k \ell} + \delta_{ik} \delta_{j \ell} + \delta_{i\ell} \delta_{jk} )$.

\section{Nematic Order and Surface Curvature}

\label{curvature}

Here, we describe the relationship between interfacial shape (surface curvature) and the nematic order parameter $Q^\alpha_{ij}({\bf x})$. We adopt the limiting form of the interfacial order parameter in the strong segregation limit, main text eq. (5), which can be rewritten as
\begin{equation}
\label{Seq:interJij}
Q^{\alpha}_{ij} ({\rm inter.}) = \frac{a^2}{60} \phi_\alpha \big[ \partial_i \partial_j \ln \phi_\alpha - \frac{ \delta_{ij} }{3} \nabla^2 \ln \phi_\alpha \big] .
\end{equation}
We can relate the structure of the order parameter to the shape of isosurfaces, level sets of constant $\phi_\alpha$.   Introducing orthonormal coordinate directions $\{ {\bf e}_1,  {\bf e}_2, {\bf e}_3 \}$, where ${\bf e}_3 = {\bf N}$ is the isosurface normal, ${\bf N} = \nabla \phi_\alpha/| \nabla \phi_\alpha|$ and ${\bf e}_1$ and ${\bf e}_2$ span the tangent plane of the surface (i.e. ${\bf e}_1 \times {\bf e}_2 = {\bf N}$), we project the nematic tensor into this tangent basis, $Q^\alpha_{IJ} = ({\bf e}_I)_i Q^\alpha_{ij} ({\bf e}_J)_i$ where indices $I, J, K$ refer to the surface bases.   

Let us represent the full form of $j_{IJ} \equiv ({\bf e}_I)_i ({\bf e}_J)_i \partial_i \partial_j \ln \phi_\alpha$ as the $3\times 3$ symmetric matrix,
\begin{equation}  \label{Seq:Jmat}
{\bf j} \equiv \left(\begin{array}{ccc}
\mathcal{K}_{11}& \mathcal{K}_{12}& \mathcal{F}_{1}\\
\mathcal{K}_{21}& \mathcal{K}_{22}& \mathcal{F}_{2}\\
\mathcal{F}_{1}& \mathcal{F}_{2}& \mathcal{F}_{3}\\
\end{array}\right)
\end{equation}
where $\mathcal{K}_{IJ}$ and $\mathcal{F}_{I}$ are in-plane and normal components, respectively.  From the definition of surface normal we have $\nabla \ln \phi_\alpha = v_\alpha {\bf N}$ where $v_\alpha =  \phi_\alpha^{-1} \partial_N  \phi_\alpha$ is a scalar function. The in-plane components of ${\bf j}$ become
\begin{equation}
\mathcal{K}_{IJ} = {\bf e}_J \cdot \big[  ({\bf e}_I \cdot \nabla) v_\alpha {\bf N} \big] = v_\alpha  {\bf e}_J \cdot \big[  ({\bf e}_I \cdot \nabla) {\bf N} \big] \ \ \ \ {\rm for} \ I,J = 1,2 , 
\end{equation}
where we used the fact that ${\bf e}_I \cdot {\bf N}=0$ for $I=1,2$.  Using $ {\bf e}_J \cdot \big[  ({\bf e}_I \cdot \nabla) {\bf N} \big] =-{\bf N}_J \cdot \big[  ({\bf e}_I \cdot \nabla) {\bf e}_J \big] = - C_{IJ}$
\begin{equation}
{\cal K}_{IJ} = -v_\alpha C_{IJ} ,
\end{equation}
where $C_{IJ}$ is the 2D curvature tensor of the surface~\cite{Kamien}.
 
Similarly, since $ {\bf N} \cdot( \partial_i  {\bf N} )=0$
\begin{equation}
{\cal F}_I = j_{I3}=j_{3I}={\bf N} \cdot \big[ ({\bf e}_I \cdot \nabla) v_\alpha {\bf N} \big] = ({\bf e}_I \cdot \nabla) v_\alpha  .
\end{equation}
Using the definition of $j_{IJ}$ we recover the form of eq. (6) from,
\begin{equation}
Q^{\alpha}_{IJ} ({\rm inter.}) = \frac{a^2V}{60N} \phi_\alpha \big[ j_{IJ} - \frac{ \delta_{IJ}}{3} j_{KK} \big].
\end{equation}
Note that since $v_\alpha \approx \Delta^{-1}$, the in-plane components of ${\cal F}_I $ may be viewed as variation of interfacial thickness along the isosurface.  Thus when interfacial thickness is approximately constant in-plane (or $F_{1} \sim F_{2} \ll F_3$), then $j_{IJ}$ and $Q_{IJ}$ are diagonal in the basis spanned by the principle curvature axes in-plane (for which $C_{12}=C_{21}=0$) and interface normal.

\begin{figure}[t]
\includegraphics[width=0.9\textwidth]{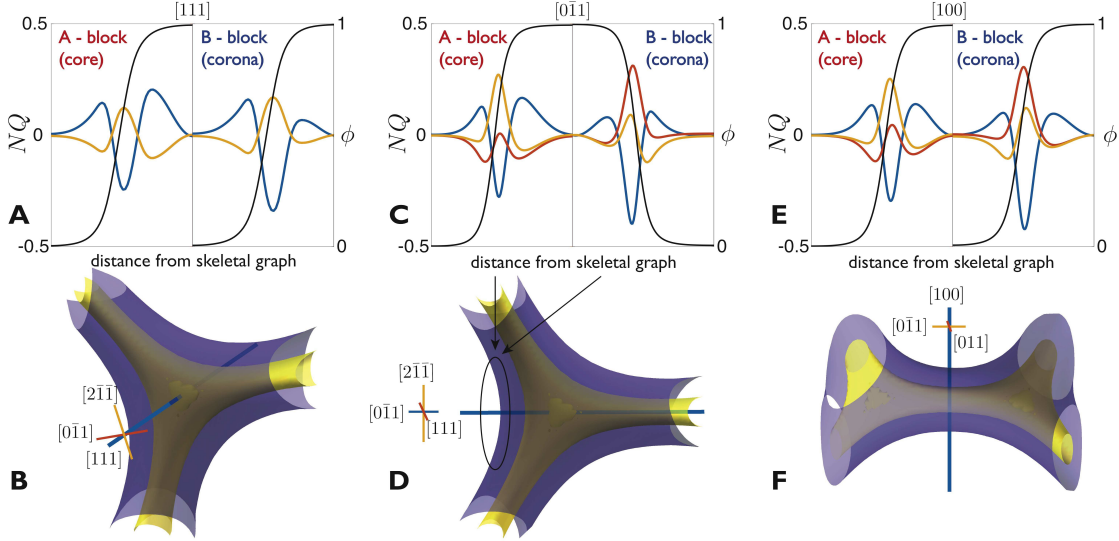}
\caption{Nematic order parameter profiles of minority (left panel) and majority (right panel) domains of gyroid morphology plotted as a function of distance from the underlying skeletal graph. Black curves represent density profiles. Projection of nematic OP along (A-B) [111] (blue) direction passing through centre of the node, overlapping yellow and red curves represent order along tangential directions and show that OP is uniaxial through the node; (C-D) projection along [0$\bar{1}$1] direction passing through the centre of a strut, profiles in C are plotted for interfacial region highlighted in D). Highly dissimilar peak values of interfacial orientation in tangential principal curvature directions (red and yellow curves) stems from opposite sign of principal curvature;  (E-F) Similar profiles along [100] direction passing through the tangential plane at centre of the strut.  Minor domain geometry is shown in (B), (D) and (F) through composition isosurfaces at $\phi_A= 0.85$ (blue) and $\phi_A=0.93$ (yellow).}
\label{fig:S3.1}
\end{figure}

\section{Nematic Order in the Double Gyroid Phase}

Here, we describe the nematic order parameter characterization of the double gyroid phase.  The minority domains of gyroid phases are formed from 2-interconnected networks, each is composed tubular domains meeting a cubic array of three-fold coordinated junctions.   The majority blocks form the ``matrix" domain between the two networks, the mid-surface of which is the Gyroid minimal surface~\cite{schick}.  Due to the variation of local interface shape throughout the morphology, both magnitudes and principle directions of nematic order that vary throughout.  Specifically the local curvatures vary considerable, leading to a range of segment order along different directions.  While the nematic director (associated with the largest eigenvalue of $Q^A_{ij}$) at the DG surface is depicted in the main text, here we analyze nematic ordering in the ``core" and ``coronal" brush domains of the phase along different directions representing maximal and minimal curvature (Gaussian is more variable than mean curvature on the DG interface~\cite{matsen_bates}).  

In Fig.~\ref{fig:S3.1} we show profiles of the nematic order parameter along 3 symmetry axes of a gyroid at $\chi N =30$ and $f=0.33$.  Profiles show projections of the nematic order parameter onto orthogonal basis vectors that pass through the centre of the 3-fold junction [111] (Fig.~\ref{fig:S3.1}A,B), along the tubular ``strut" joining two junctions [110] (Fig.~\ref{fig:S3.1}C,D) and perpendicular to the strut at its centre [100] (Fig.~\ref{fig:S3.1} E,F). Consistent with other morphologies, we find projection of the nematic order parameter normal to the interface to be negative at the interface. Lack of rotational symmetry at the interface leads to dissimilar ordering in orthogonal tangential directions, with segments favoring to orient along most negatively curved directions. Majority blocks have larger interfacial orientation but weaker normal, brush-like orientation than minority domain block segments. Of the three directions, minority block segment extension normal to the interface is strongest along the [111] direction that passes through the 3-fold axes due to the locally flattening of the interface along that direction.  In contrast, the highest (negative) surface curvature occurs along the [100] axes that bisects the tubular struct, leading to the smallest normal component $Q_{NN}^A$ in the (core) brush domain.

\begin{figure}
\includegraphics[width=0.9\textwidth]{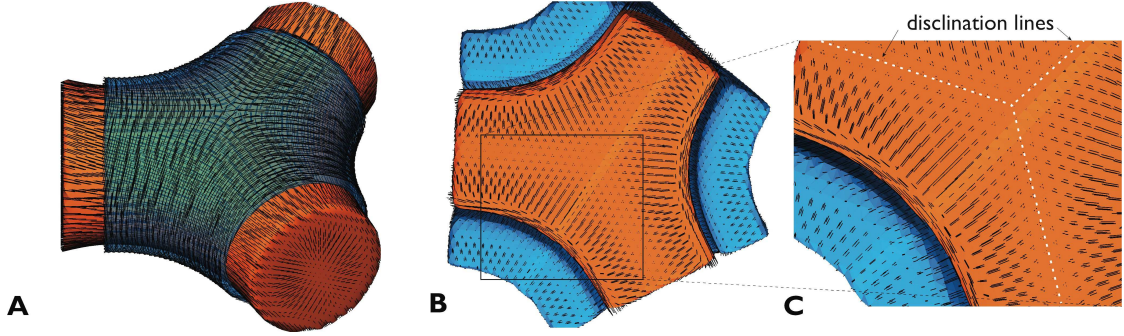}
\caption{(A) Orthogonal segment orientation at the IMDS (or $\phi=1/2$), with director of the core and coronal block segments shown on the respective orange and blue surfaces ; (B-C) Segment orientation normal to the interface in the ``brush" regions of both major (orange) and minor domains (blue).}
\label{fig:S3.2}
\end{figure}

Fig.~\ref{fig:S3.2} illustrates features of the majority block texture the forms between the two (minor) tubular domains.   Fig.~\ref{fig:S3.2}A shows the nematic director of the major block segments at the interface (blue surface) to be perpendicular to the director of the minor block segments (orange surface).  In Fig.~\ref{fig:S3.2}B, a cross section of the composite core/coronal phases of the DG is shown to highlight the normal orientation of the both brush domains away from the surface (the outer domain extends to roughly the location of the Gyroid minimal surface).

\section{MD simulation and Order Parameters analysis}

\subsection{Simulation Details}
For both lamellar and cylindrical phases, we use a simple Kremer-Grest bead-spring model~\cite{Grest1996} where the following finitely extensible nonlinear elastic (FENE) and fully repulsive Lennard-Jones (LJ) potentials are used for bonds and all pairwise interactions, respectively. 
\begin{equation}
 U_{\mathrm{FENE}}(\mathrm{r}) = -0.5kR_{0}^2\mathrm{ln}\Big ( 1-\frac{r^2}{R_0^2} \Big )
\end{equation}
where a spring constant of $k=30\epsilon/\sigma^2$ and maximum length of $R_0=1.5\sigma$ are used to appropriately avoid chain crossing or breaking.
\begin{equation}
 U_{\mathrm{LJ}}(\mathrm{r}) = \begin{cases}
 4\epsilon_{ij}\Big [ \Big (\frac{\sigma_{ij}}{r} \Big )^{12} - \Big (\frac{\sigma_{ij}}{r} \Big )^{6} + \frac{1}{4} \Big ] &\mbox{if } r\leq r_c \\
0  &\mbox{if } r > r_c 
 \end{cases}
 \end{equation}
which is the standard LJ potential cut off and shifted to 0 at $r_c=2^{1/6}\sigma$. $\epsilon_{ij}$ is the interaction strength between monomers $i$ and $j$, which is equal for like monomers ($\epsilon_{AA}=\epsilon_{BB}=\epsilon$ ) but increased for unlike monomers ($\epsilon_{AB}>\epsilon$ ) to match with considered $\chi$ values. $\sigma$ is the length scale of interaction between $i$ and $j$ monomers, and all monomer sizes equal ($\sigma_{AA}=\sigma_{BB}=\sigma_{AB}=\sigma$). All monomers have unit mass (1.0m). 

The Flory-Huggins interaction parameter $\chi$ is mapped to $\epsilon_{AB}$ by the numerical integration of
\begin{equation}
\chi=\Big (\epsilon_{AB}-\epsilon \Big ) \frac{4\rho}{k_B T} \int_0^{r_C} \! \Big [ \Big (\frac{\sigma_{ij}}{r} \Big )^{12} - \Big (\frac{\sigma_{ij}}{r} \Big )^{6} + \frac{1}{4} \Big ] g(r) \mathrm{d}r
\end{equation}
where $k_B$ is the Boltzmann constant and $\rho$ is the monomer number density. The intermolecular $g(r)$ is directly obtained from a simulation of homopolymer melt with chain length $N$ of 100 beads. The first-order approximation thus obtained is $\chi=0.66(\epsilon_{AB}-\epsilon)/k_B T$.

\begin{figure}
\includegraphics[width=0.7 \textwidth]{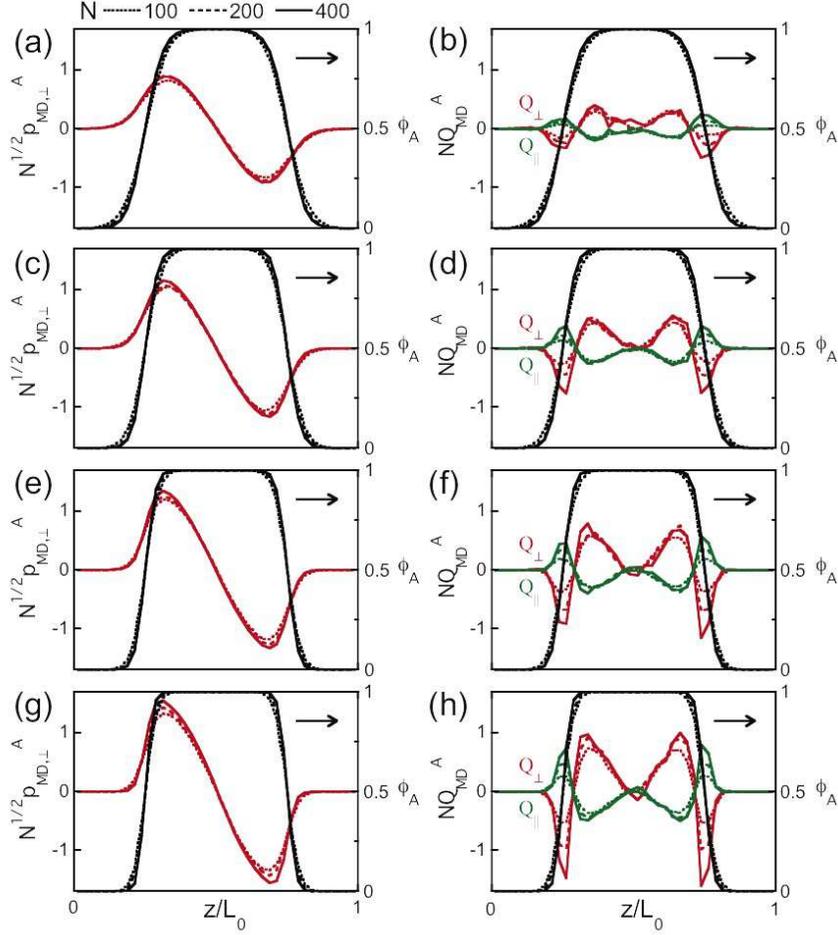}
\caption{ Scaled compositional, vector, and tensor order parameters for A monomers of lamellae for three different chain lengths ($N$ = 100, 200, 400) at $\chi N$ = (a,b) 50, (c,d) 80, (e,f) 120, and (g,h) 200}
\label{fig:S4.1}
\end{figure} 

A Langevin thermostat with damping parameter 1.0$\tau$ (the reduced unit of time $\tau=\sigma(m/\epsilon)^{1/2}$) is used to keep the reduced temperature $T=\epsilon/k_B$. Simulation is performed in the open source molecular dynamics (MD) package with timestep 0.0115$\tau$ and periodic boundary conditions. 

We create linear AB diblock chains with fraction of A monomer being 0.5 for lamellae and 0.25 for cylinders. The simulations of both structures are started from initially phase separated configurations. The initialization of lamellae follows the procedure described in ref. ~\cite{seo}.  We have 1600 polymers equally distributed in 4 lamellar layers for three different chain lengths, N=100, 200, and 400 each of which is simulated at 4 $\chi N$ values of 50, 80, 120, 200. For cylindrical structures, we use a similar approach of initialization of random walks on either side of the cylindrical interface, creating 12 cylinders parallel to the $z$ axis (the cross section contains 3 repeats of the shorter side of the rectangular unit cell in the $x$ direction and 2 of the longer sides in $y$) where each cylinder contains A monomers of 200 polymers. The initial total monomer number density and interfacial coverage density for cylinders are 0.85$\sigma^{-3}$ and 0.12$\sigma^{-3}$, respectively. We consider only one system at $\chi N$ =200 with $N=100$ for the cylindrical structure. The table below shows values used in our simulations for each condition of $\chi N$ and $N$.

\begin{table}[ht]
\begin{tabular}{|c|c|c|c|}\hline $\chi {\rm N}$ & N=100 & N=200 & N=400 \\\hline 50 & 1.7576 & 1.3788 & 1.1894 \\\hline 80 & 2.2121 & 1.6061 & 1.3030 \\\hline 120 & 2.8182 & 1.9091 & 1.4545 \\\hline 200 & 4.0303 & 2.5152 & 1.7576 \\\hline 
\end{tabular}
\label{Parameters used in MD simulations.}
\end{table}

Before switching on the intermolecular LJ interactions, monomers that may be overlapping are pushed off of each other using a soft potential $U_s(r)=A\Big [ 1+cos(\pi r/r_c) \Big ]$ for $r<r_c$ , where $r_c=2^{1/6}\sigma$ and $A$ linearly increases from 0 to 250 for 57.5$\tau$. Then the equilibration is done using a Nos\'e-Hoover barostat to keep the pressure 5$\epsilon\sigma^{-3}$, which gives the density approximately 0.85$\sigma^{-3}$ for analogous homopolymer melts, with a damping parameter of 10$\tau$. The $x$ and $y$ box lengths are  constrained to be equal (barostatting in those directions is coupled) for lamellae, while the $z$ direction can adjust to create the proper lamellar domain spacing. The box lengths are not constrained in any direction for cylinders. During the equilibration of lamellar systems of $N=400$, a double-bridging algorithm is employed to allow chain crossing (and faster equilibration), wherein close enough bonds are allowed to swap under the Boltzmann acceptance criterion~\cite{Auhl2003}. The bond swapping criterion is applied for two bonds on the same site from different polymers (the bonds are labeled from 1 to $N-1$ from the end of A blocks to the end of B blocks) if their distance is less than 1.3$\sigma$, and such test occurs for 50\% of bond pairs every 0.115$\tau$. All systems are equilibrated for 345,000$\tau$ and during this time, box sizes, total number density, polymers' mean-squared end-to-end distance $\langle R_{ee}^2\rangle$, and polymers' mean-squared radius of gyration $\langle R_g^2\rangle$ are monitored. The initial microphase separated structures are conserved throughout our equilibration time and all of the monitored structural properties settle to be close to their final values within 57,500$\tau$ for $N=100$ and 230,000$\tau$ for $N=200$, 400 for all $\chi N$ values

\subsection{Analysis of Order Parameters}

The compositional order parameter for A monomers is defined in MD as follows: \\
\begin{equation}
\rho^A_{MD}(z \; {\rm or} \; r) = \frac{\sum_\alpha^{N_A}  \rho_{\hat{r}_\alpha}(z \; {\rm or} \; r)}{\sum_\alpha^{all}\rho_{\hat{r}_\alpha}(z \; {\rm or} \; r)}
\end{equation}
where $z$ is for lamellae binned in $z$ direction whereas $r$ is for cylinders binned in radial direction from the center of cylinders. We have 40 bins across a lamellar layer and 20 bins for a cylinder from the center to the shorter length of the cross section's rectangular unit cell of the cross section (D). We collected radially averaged data for cylinders for improved statistics (rather than, for instance, considering a narrow region along the shortest lines connecting nearest neighbor cylinders); note that data is not collected in the interstitial regions further than D from any cylinder center  where $\hat{r}_{\alpha}$ is an $\alpha$-$\alpha$ bond vector ($\alpha$ is A- or B-type bead) and defined to direct inward to the AB bonds from each end of polymers and  is the number density of  in the bin. For lamellae, coordinates are relative to an averaged center of mass point of the A layers obtained by conceptually overlapping the 4 layers in the box. For cylinders,  the coordinates are relative to an averaged center of mass point is calculated separately for each cylinder that is distinguished by a cluster analysis. All of the order parameter profiles as a function of r are averaged for the 12 cylinders. We note that polymers do not cross from one cylinder to the other during our simulation time, which is not surprising for this high value of $\chi N$ (200), thus the cluster analysis does not have to be updated for each configurations.

\begin{figure}
\includegraphics[width=0.65 \textwidth]{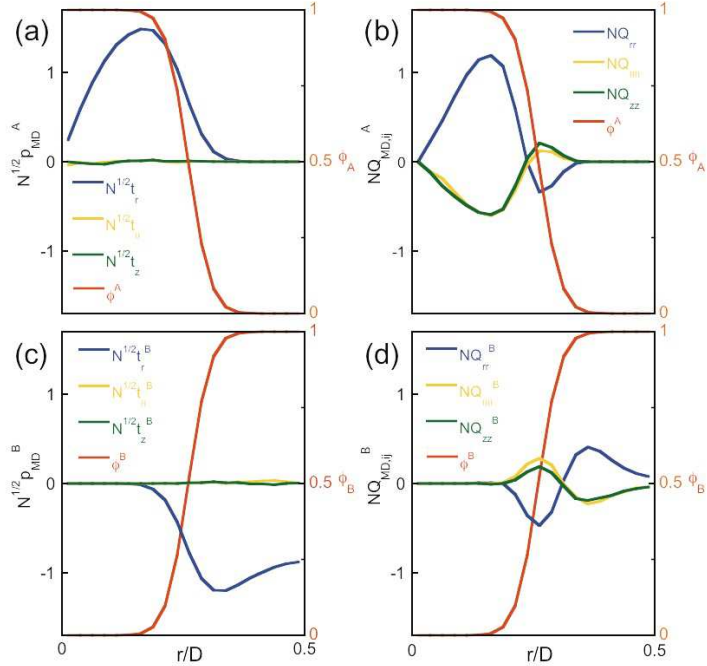}
\caption{ Scaled compositional, vector, and tensor order parameters for (a,b) A and (c,d) B for $N$ = 100 and $\chi$ = 2}
\label{fig:S4.2}
\end{figure} 

The polar (vector) order parameter is 
\begin{equation}
p^A_{MD}(z \; {\rm or} \; r) = \frac{\sum_\alpha^{N_A}\hat{r}_\alpha / V_{z \; {\rm or} \; r}}{\sum_\alpha^{all}\rho_{\hat{r}_\alpha}(z \; {\rm or} \; r)}
\end{equation}
where  $V_{z \; \mathrm{or} \ r}$is the bin volume. For lamellae, the perpendicular vector order parameter $p^A_{MD,\perp}$ is obtained by projecting the bond vectors on the unit vector of $z$ axis, or only considering the z coordinates of the bond vectors. For cylinders, we project the bond vectors onto three different vectors one of which is perpendicular to the cylinder ($p^A_r$ is the radial vector connecting the center of cylinder to the midpoint of a bond) and two of which are parallel to the cylinder ($p^A_z$, the unit vector of $z$ axis and $p^A_{\theta}$, tangential vector obtained from the radial vector).
The nematic (tensor) order parameter is 
\begin{equation}
Q^A_{MD,ij}(z \; {\rm or} \; r) = \frac{\sum_\alpha^{N_A}[(\hat{r}_\alpha)_i (\hat{r}_\alpha)_j - \delta_{ij}/3]/V_{z \; {\rm or} \; r}}{\sum_\alpha^{all}\rho_{\hat{r}_\alpha}(z \; {\rm or} \; r)}
\end{equation}
where $\delta_{ij}$ is the Kronecker delta and $i$, $j$ are x,y,z. For lamellae, $Q^A_{MD,\perp}=Q^A_{MD,zz}$ and $Q^A_{MD,\parallel}=(Q^A_{MD,xx}+Q^A_{MD,yy})/2$. The configurations were saved every 115 steps during the equilibration time for all systems. both compositional and vector order parameters are averaged for the last 100 configurations while tensor order parameters are averaged for the last 400 to further improve the statistics.

\bibliography{OPpaper}

%
%
%
%

\end{document}